\documentclass[aps,pra, onecolumn,14 pts,a4paper,showpacs]{revtex4}

\usepackage{amssymb,amsmath,amsfonts,graphicx}
\usepackage{bm}

\begin{document}

\title{The transition to turbulence in parallel flows: transition to turbulence or to regular structures} 

\author{Yves Pomeau$^1$ and Martine Le Berre$^2$. }
\affiliation{$^1$ Department of Mathematics, University of Arizona, Tucson, AZ 85721, USA.
\\$^3$ Institut des Sciences Mol\'eculaires d'Orsay ISMO
- CNRS, Universit\'e Paris-Sud, Bat. 210, 91405 Orsay Cedex, France.
} 
\date{\today}

\begin{abstract}
\textbf{Abstract}

We propose a scenario for the formation of localized turbulent spots in transition flows, which is known as resulting from the subcritical character of the transition. We show that it is not necessary to add ''by hand'' a term of random noise in the equations, in order to describe the existence of long wavelength fluctuations as soon as the bifurcated  state is beyond the Benjamin-Feir instability threshold.
We derive the instability threshold for generalized complex Ginzburg-Landau equation which displays subcriticality. Beyond and close to the Benjamin-Feir threshold we show that the dynamics is mainly driven  by the phase of the complex amplitude which obeys Kuramoto-Sivashinsky equation while the  fluctuations of the modulus are smaller and  slaved to the phase (as already proved for the supercritical case). On the opposite, below the Benjamin-Feir instability threshold, the bifurcated state does loose the randomness associated to turbulence so that the transition becomes of the mean-field type as in noiseless reaction-diffusion systems and leads to pulse-like patterns. 

\end{abstract}

\maketitle

\section{Outline}

We outline the main ideas behind the present understanding of the transition to turbulence in parallel flows, hopefully within reach of a reader unfamiliar with fluid mechanics.
The transition to turbulence in parallel flows is notoriously
difficult, specifically we notice that most textbooks treat  this
subject via  linear stability which actually does not help much to
understand phenomena which are non linear including at the onset. Nevertheless some understanding of the observed  transition can be derived from an in depth analysis of the weakly nonlinear analysis (amplitude equations) in the pertinent regime. Following the example of Landau, who introduced the idea of amplitude equation without giving a hint of the way it can be derived from the fluid equations, we leave out the intricacies of the derivation of the amplitude theory from the original Navier-Stokes equations, an approach that has the big advantage to not bury the reader under a stack of algebraic formulae helping very little to understand the final result. In this approach the space dependent behavior can be understood thanks to a "generic" extension of the original Landau's model for amplitude equations to include a slow space dependence for linearly unstable waves. The main result of our analysis is that noise may be or may not be present in the flow structures that develop spontaneously inside the localized turbulent spots. 

 That turbulence can remain localized in transition flows have long been a mystery. Now it is well understood that this is because the bifurcation to turbulence is subcritical. In a range of parameters there are two possible states, the laminar and the turbulent state. The turbulent spots are domains where the fluid motion is in "the other state" existing at the given Reynolds number besides the laminar flow. This explanation, based on concepts borrowed from the dynamics of systems with a potential energy, does not take into account that the turbulent state is not steady but filled of random fluctuations, with the consequence that the transition is of the directed percolation type, at least in some cases.

We begin with a rather general discussion of the theory of the transition to turbulence in parallel flows. Even though in the present case linear stability theory 
 is  fairly complex (linked to the fourth order Orr-Sommerfeld equations,  the topic of Heisenberg PhD thesis \cite{heisenberg}), unfortunately it brings  very little help for understanding the actual transition which relies on a fully nonlinear process. 
We sketch first the analysis of the saddle-node transition, the kind of transition occurring in parallel flows, which are linearly stable in the range where transition is observed. It may even happen, for instance in circular Poiseuille flow, that the flow is linearly stable for {\it{any}} value of the Reynolds number. We recall Landau's theory of bifurcation, concluding that it is supercritical or subcritical according to the sign of a coefficient in the so-called Landau expansion, a case of Poincar\'e normal form. When the transition is subcritical we explain how to extend Landau's equation to include a possible space dependence of the amplitude of the fluctuations off the laminar state. This leads quite naturally to the property of localized states, growing or collapsing according to the value of  various parameters. 

In this framework, the transition is not necessarily from laminar to turbulent localized states, because the "new" state occurring at the bifurcation is a plane wave of finite amplitude, {\it{a priori}} quite different of what is understood as turbulent because the randomness in space and time associated to turbulence is not a property of such a finite amplitude wave. Such randomness is a consequence of the Benjamin-Feir instability of the phase of finite amplitude waves. Finite amplitude waves may be either stable (no randomness then) or unstable against long wavelength modulations of their phase: this is the Benjamin-Feir instability. We explain next that, once nonlinear effects are included, this instability yields the Kuramoto-Sivashinsky equation for the fluctuations of the phase, an equation known for having turbulent solutions in large enough space domains. Somehow this yields a complete picture of the transition to turbulence: localized turbulent domains are explained by the subcritical character of the transition and the turbulence inside follows from the  Benjamin-Feir  instability of  finite amplitude waves in the turbulent spots. 

The originality of the present work lies  in the fact that it concerns a situation not yet studied:  \textit{phase turbulence associated to a sub-critical bifurcation}. The concept of phase turbulence  has been introduced by Kuramoto \cite{Kuramoto}  to describe the chaotic state of many identical self-oscillators interacting by diffusive coupling, without topological defect of the phase field. Later  Manneville and Chat\'e described phase turbulence of a continuous complex field in the framework  of complex cubic Ginzburg-Landau equation (CGL) \cite{Manneville-Chate}  in the super-critical case. In their study (and other works afterwards)  cubic nonlinearity is enough to describe phase turbulence which occurs close to Benjamin-Feir threshold. In those studies turbulence arising from the Kuramoto-Sivashinsky instability is spread over the whole space because it exists only when the trivial homogeneous state is linearly unstable. 
Here the sub-critical property which requires to include at least a quintic nonlinear term, is essential because our aim is to describe turbulent spots co-existing with a laminar featureless flow, at least transiently. 
  
We note that up to now  quintic CGL  solutions have been investigated in the subcritical domain, but only  below the Benjamin-Feir threshold, where turbulent states are not expected. In those studies pulse-like patterns have been predicted and experimentally observed : the pulses are formed via fronts joining the two stable homogeneous solutions  \cite {quintic-pulse}.

\section{From Reynolds to Landau: subcritical vs supercritical bifurcation}

Beginning with Reynolds many experiments have put in evidence that parallel flows bifurcate to turbulence via a regime such that turbulence or more generally the bifurcated state remains localized in well separated domains having received various names. In careful studies Emmons \cite{Emmons} showed that, above  a range of Reynolds number, a Blasius boundary layer displays what are called now "Emmons spots" growing with a well defined arrowhead shape surrounded by laminar flow. The inside of Emmons spots shows up a quite regular and reproducible structure of rolls with axis along the stream-wise direction, making them a good candidate for Benjamin-Feir stable finite amplitude waves in the bifurcated state.  

The observation of localized structures with a turbulent inside is explained by the fact that the transition is subcritical, namely that the jump to the turbulent/bifurcated state (we shall denote later this new state as "turbulent" although we shall argue that it is not necessarily turbulent because perturbations there may be chaotic or not) grow out of finite amplitude initial fluctuations. Moreover the turbulent domain does not grow at an exponential rate but at a finite and well defined velocity of expansion.  Because this is directed to a non specialists audience and readership, we skipped the feedback between the mean flow and the fluctuations of finite amplitude. In real life, the growth of the fluctuations changes at least locally the mean flow, basically because turbulence acts like a local increase of an efficient viscosity. Such an increase of the effective viscosity changes the flow and so the conditions for the development of the instability. This poses rather complex question that have been dealt with in some particular cases. We refer the interested reader to the original publications  \cite{feedback} on this feedback effect. 

This property of contamination of one state by the other replaces the standard linear instability in the case of subcritical bifurcation: the amplitude equation written below with a space dependent term (Equation (\ref{eq:LandQfull2cc}) for the most general form) has, in the case of a subcritical bifurcation, two  homogeneous states {\it{linearly stable in an average sense}}: the laminar and the "turbulent" one (see below for what we mean by "linear stability" of a turbulent state). If one considers as an initial condition for this equation a solution existing in a half space and the other one in the other half, this solution will evolve toward a solution asymptotics at the two infinities of the two different states, the laminar and the bifurcated one, but one state will invade the other at constant speed and the boundary will move at constant speed. This replaces the standard concept of growth of linearly unstable fluctuations by the one of growth (or decay) of a state by contamination of the other:  which one contaminates the other depends of course of the sign of the velocity of the front separating the two states.  This sign can be decided without solving the equation for the front whenever there is a potential formulation of the problem: the state with the lowest energy tends to invade the one with the highest energy, but in general- as for instance in the model of Eq.(\ref{eq:LandQfull2cc}) with complex coefficients the direction of the front motion can be decided only by solving the equation for it. Suppose we have two steady state solutions of the amplitude equation, one without perturbation to the base flow, namely with amplitude $A = 0$, the other with a finite amplitude depending eventually periodically on time like $ A = A_0 e^{i\omega t}$. Then the front solution is of the form $A_f (x - v t) e^{i\omega t}$, $x$ stream-wise coordinate and $v$ front velocity. This velocity $v$ is derived by putting the function $A_f$ into a model like Eq.(\ref{eq:LandQfull2cc}) and by imposing that $A_f$ tends to zero when its argument tends to $\pm\infty$ and tends to $A_0$ as its argument tends to  $-\pm\infty$. This velocity is a nonlinear eigenvalue necessary to cancel an exponentially growing mode with respect to the variable $(x - v t)$ near $A = A_0 e^{i\omega t}$. 
 
This agrees with the experimental results and also with the simplest model of a gradient flow \cite{gradflow} for the amplitude of fluctuations with a diffusion term. This model belongs to the class of the so-called reaction-diffusion systems. This simple analytical picture does not follow in general from a systematic derivation based on the fluid equation, the Navier-Stokes equations in the present case. One of the many reasons (although this is not that obvious - see below) for this apparent lack of connection with the basic equations is because the reaction-diffusion equations, at least for a single species, have the property of being a gradient flow, which is well-known not to be the case for the Navier-Stokes equations. However things are not that obvious because the amplitude equations, the ones of our theory, have formally such a structure in the absence of a diffusion term. However once the diffusion term is included, the amplitude equation includes complex coefficients so that, in general, it cannot be derived from a real potential energy. Therefore it can describe a permanently turbulent, namely time dependent,  state.

The idea presented below can be summarized as follows: we derive first the general form of Landau's amplitude equation from general symmetry argument, but, contrary to what Landau did, for the amplitude itself and not for the intensity (the modulus square of the amplitude).  The result is an equationn including two "arbitrary" functions of the modulus of the waves denoted as $f(.)$ and $g(.)$ which can be derived  in principle numerically from the study of wave solution of Navier-Stokes equations with finite amplitude. 

\subsection{Time dependent perturbations}

 Without a space derivative, the amplitude equation can be reduced to Landau's equation for the intensity, plus another equation for the phase whose solution can be found once the equation for the intensity is solved. This approach poses three questions:

1) Obviously it cannot describe the occurrence of localized structure, because it does not include any space dependence.

2) the nonlinear terms are limited to the fifth power of the amplitude, which requires to explain why the seventh order and higher order effect are neglected.  

3) Why limit the expansion to a	 finite order and not replace truncated sums by functions?  

The physical situation represented by Landau's amplitude equation is not that common. It could be for instance the transition observed in the wake of a sphere. This wake shows a bifurcation from steady to periodic oscillations at a Reynolds number around 40. Close to the transition one can assume that the oscillating part of the wake adds to the fluid velocity a small contribution depending on space and time like $A'(t) F(\bf{r})$ where $F(\bf{r})$ is a function of position which can be derived in principle from the Navier-Stokes equations by perturbation near the solution for a steady flow. Because they are oscillations one can write the time dependent function $A'(t)$ as the sum of a complex function $A$ and its conjugate. Those two functions are solutions of one equation and its complex conjugate. Near the onset of oscillations, following Landau, one assumes that this equation can be expanded in powers of  the small amplitude $A$. This yields the following Poincar\'e normal form of the equation. 


Let us consider point number 2 above and write the equation for the amplitude up to the fifth order term:
\begin{equation}
\frac{d A }{dt}  =  (\delta_R + i \delta_I ) A  + (\beta_R + i \beta_I) |A|^2 A - (\gamma_R + i \gamma_I ) |A|^4 A 
\mathrm{.}
\label{eq:LandQ}
\end{equation}
 In this equation the amplitude $A$ is a complex function of time $t$. 
The symbols $\beta_R$, $\gamma_R$, $\delta_R$ etc. are for real numbers which can be computed from the Navier-Stokes equation at least in principle for a given instability, although this can be very cumbersome, particularly for $\gamma_R$ and $\gamma_I$. Notice that this equation is invariant under multiplication of $A$ by an arbitrary phase factor $e^{i\varphi}$. This reflects the property of invariance of the solution under an arbitrary translation in time, a consequence itself of the property that the perturbation we consider is a perturbation of a steady solution of the Navier-Stokes equations. 

Multiplying $A$ by $e^{i \delta_I t }$, one can get rid of the term proportional to $\delta_I$ in the equation. This yields (without introducing a new symbol for $A e^{i \delta_I t }$):
 \begin{equation}
\frac{d A }{dt}  =  \delta_R A  + (\beta_R + i \beta_I) |A|^2 A - (\gamma_R + i \gamma_I ) |A|^4 A 
\mathrm{.}
\label{eq:LandQred}
\end{equation}

Landau's amplitude equation in its original meaning is derived by multiplying Eq.(\ref{eq:LandQred}) by the complex conjugate $A^*$, to obtain:

 \begin{equation}
\frac{d I }{dt}  =  \delta_R I  + \beta_R I^2 - \gamma_R I^3
\mathrm{.}
\label{eq:LandQredint}
\end{equation}
where $I = |A|^2$. This equation is purely real and has the variational (or gradient flow) structure alluded to before, because it can be written as 

 \begin{equation}
\frac{d I }{dt}  = -  \frac{\partial U}{\partial I}
\mathrm{.}
\label{eq:LandQredintvar}
\end{equation}
with 
 \begin{equation}
 U =  - \frac{\delta_R I^2}{2 } -  \frac{\beta_R I^3}{3} + \frac{\gamma_R I^4}{4} 
 \mathrm{.}
 \label{eq:pot}
\end{equation}

Therefore any solution of Eq.(\ref{eq:LandQredintvar}) tends to bring down the potential $U(.)$, which makes sense if this potential is bounded from below, that is if $\gamma_R$ is positive.  The right-hand side of equation.(\ref{eq:LandQredint}) is a second degree polynomial times $I$. Therefore the steady states are either $I = 0$ or one of the two real roots, if they exist of 
 \begin{equation}
 \delta_R  +  \beta_R I - \gamma_R I^2 =0
 \mathrm{.}
\label{eq:steady1}
\end{equation}

Such roots exist if 
 \begin{equation}
  \beta_R^2 + 4  \delta_R  \gamma_R > 0
  \mathrm{.}
\label{eq:cond1}
\end{equation}

\begin{figure}
\centerline{
(a) \includegraphics[height=1.5 in]{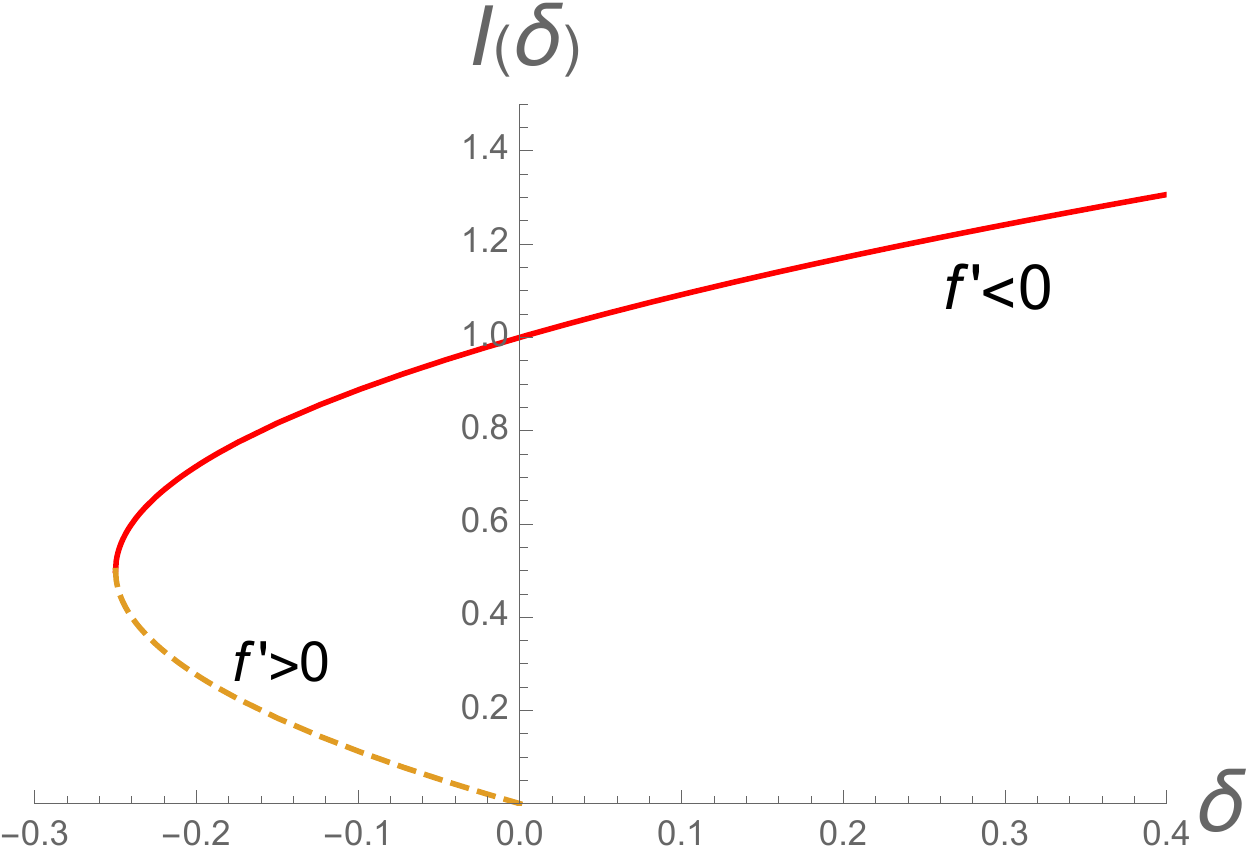}
(b)\includegraphics[height=1.5 in]{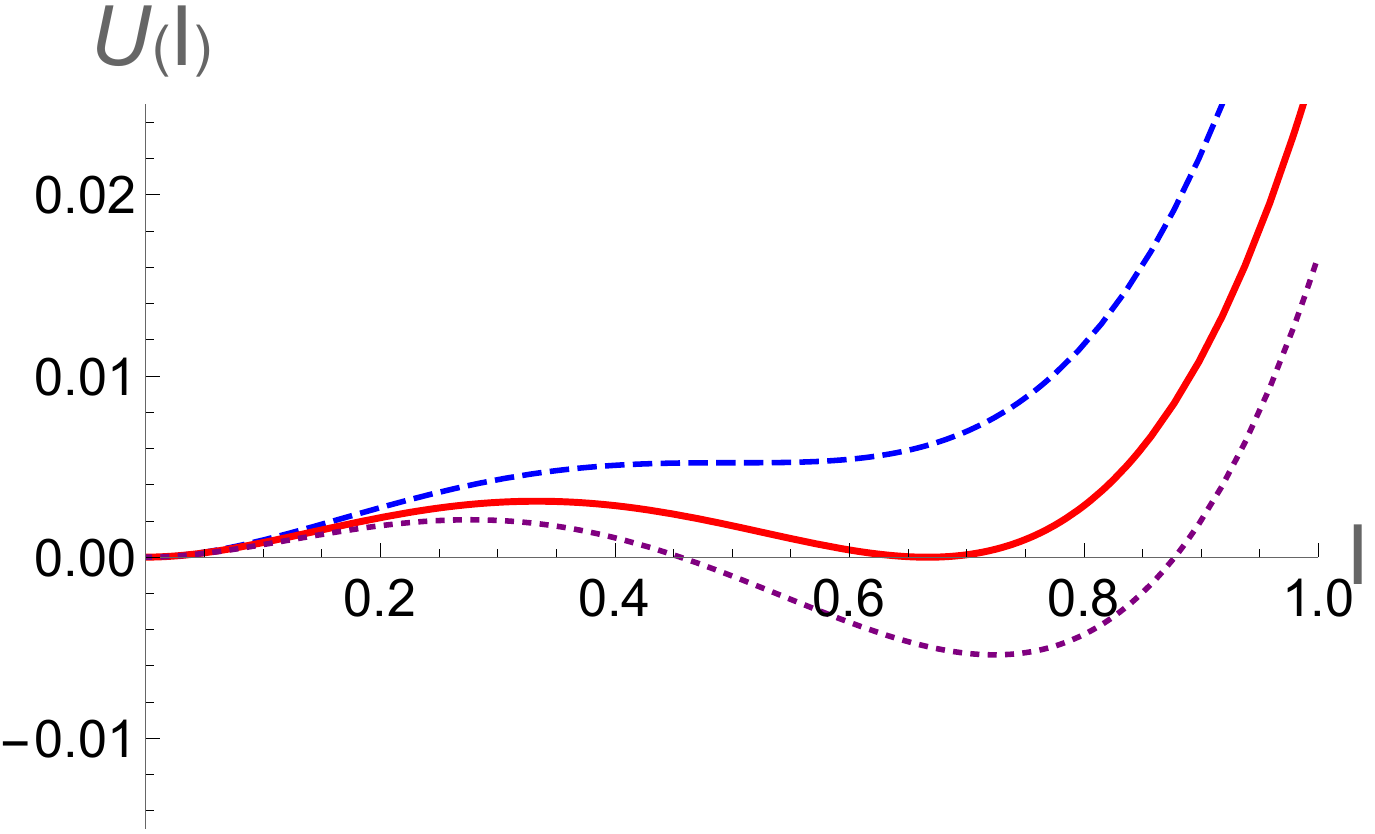}
 }
\caption{
(a) Steady state solutions of equation (\ref{eq:steady1}),
$I = |A|^2$ versus the control parameter $\delta= \delta_{R}$ .
 (b) Potentiel versus intensity, equation (\ref{eq:pot})  for $\delta= -\frac{1}{4} $ (dashed line), $\delta= -\frac{2}{9} $ (solid line) and $\delta= -\frac{1}{5} $ (dotted line), all for  $\beta_{R} =\gamma_{R}=1$.
}
\label{fig:subcr}
\end{figure}

One root corresponds to an unstable equilibrium state and the other  to a locally stable equilibrium.  This is illustrated in Fig.\ref{fig:subcr}  where we set $\beta_{R}=\gamma_{R}=1$ which can be done by rescaling  equation (\ref{eq:LandQredintvar}).  The intensity curve in Fig.(a) has a S-shape, with the unstable branch drawn in dashed line, and the stable one in full line,  corresponding  to   domains where $df(I)/dI$ positive and negative respectively (see below for the definition of f(I)).  The potential  $U(I)$,  drawn in Fig.(b) for negative $\delta $ values ( sub-critical domain),  displays $3$ extrema.  The local equilibrium $I=0$  (trivial solution, lower branch of the $S$ curve)  is the most stable one for $ -\frac{1}{4} <\delta< \delta_{M}$,    while in the other domain, $\delta >  \delta_{M}$, the upper branch  corresponds to the most stable equilibrium  state.  The middle-branch corresponds to a local maximum of the potential, then to an unstable equilibrium state, as expected. The parameter $\delta_{M}$ characterizes the Maxwell point where the two minima of the potential are  equal.  This  point  is such that  the two conditions  $\frac{\partial {U(I,\delta)}}{\partial{I}}=0 $ and $U(I, \delta)=0$ are satisfied, it gives $\delta_{M}=-\frac{2}{9}$.

Consider now the applicability of the previous theory. It relies on an amplitude equation limited to the first three terms on the right-hand side. This implies that the next order terms with respect to  the amplitude are negligible compared to the ones kept. This implies first closeness to the onset of linear instability, namely that  $\delta_R$ is small, which is sufficient to define a range of applicability of amplitude theory in the case of supercritical instabilities, because then the scale for the intensity is such that  $ I \approx   - \frac{\delta_R}{\beta_R}$. If there is linear instability $\delta_R$ is positive and, if  $\beta_R$ is positive too, there is no stable steady solution of amplitude proportional to $\delta_R$ near the instability threshold, at least unless the coefficients $\beta_R$ and $\gamma_R$ are tuned in such a way that this small amplitude steady solution exists near threshold. Looking at the second degree polynomial giving the steady state solutions, one finds that such small amplitude solution exist if $\gamma_R$ is of order one, although both $\delta_R$ and $\beta_R$ are such that $\beta_R \sim \left(\delta_R \gamma_R \right)^{1/2}$, which yields $I \sim  \left(\frac{\delta_R}{\gamma_R}\right)^{1/2}$. Because $\delta_R$ is small, this shows that, in this rather special case, the next order terms, namely the power 7, 9, etc. of $A$ in Eq.(\ref{eq:LandQred}) are negligible compared to the one kept on the right-hand side. 

Such a situation where two physical parameters ($\delta_R$ and $\beta_R$) are small requires to tune two physical quantities. Usually, the onset of linear instability is studied by taking a value of the Reynolds number, ultimately of the flow velocity in the neighborhood of this onset. But we need also to have  $\beta_R$ small to have a weakly subcritical bifurcation. In principle this could be achieved, in the case of pipe flows, by choosing a special cross section. 
Instead of this so-called quintic Ginzburg-Landau equation we shall consider a generalized form valid for perturbations of finite amplitude. It can be seen as the equation derived by carrying the amplitude expansion to all orders so that the first terms in Taylor expansion with respect  to the small amplitude $A$ are replaced by function of this amplitude representing the sum of all terms of the Taylor series. Let us write first this amplitude equation and comment it afterwards:

\begin{equation}
\frac{\partial A }{\partial t}  =    A f(|A|^2)+  i A g(|A|^2) 
\mathrm{,}
\label{eq:LandQfull2}
\end{equation}
where the function $f(.)$ and $g(.)$ are real function of the positive variable $|A|^2$. The equation for the complex conjugate reads:
\begin{equation}
\frac{\partial A^* }{\partial t}   =    A^*  f(|A|^2) -  i A^*  g(|A|^2) 
\mathrm{,}
\label{eq:LandQfull2cc}
\end{equation}
This writing of generalized amplitude equations raises two questions: what is the meaning of $A$ in those two equations and how to compute "practically" the functions $f(.)$ and $g(.)$? 

The answer to the first question bears on the amplitude expansion. This expansion relies on the assumption that there exist a wave-like solution of the Navier-Stokes equation in a parallel flow at high enough Reynolds number such that parameters like the velocity depends on space like $A(t)  e^{i{\bf{Q}}\cdot{\bf{r}}} + cc $ where ${\bf{Q}}$ is a wavenumber in the direction where the flow is infinitely extended: along the pipe in a Poiseuille flow, or in either direction of the plane in a plane Poiseuille or Couette. The computation of functions  $f(.)$ and $g(.)$ in "concrete" terms can be done numerically only. The function $f(|A|^2)$ is just a way of expressing the relationship between  the wave amplitude and its frequency. The other function, $g(|A|^2)$, is less obvious to find. it could be derived at least for small variation of the amplitude near a permanent-wave solution by looking at the exponent of stability (or instability) of this solution. 

Thanks to this writing of the amplitude equation, il will be possible to discuss in fairly general and straightforward way the Benjamin-Feir instability.

\subsection{Space and time dependent perturbations}

Returning to the main stream, we can say that the standard result of stability theory outlined above and valid for space independent amplitudes only cannot be generalized to the problem of stability of parallel flows. This is because for a parallel flow the amplitude $A$, when small, is the amplitude of a plane wave solution of the Orr-Sommerfeld equation for linear stability. This implies that, to represent physically a solution, it has to be transformed into a function depending on space, along the flow direction, like $\frac{1}{2} \left(A(t) e^{iQx} + A^*(t) e^{-iQx} \right)$ where $Q$ is the wavenumber of the most unstable perturbation and $x$ is the stream wise coordinate along the flow. This assumes that the flow is bounded in the two other dimensions perpendicular to the fluid velocity, as is the case of a pipe flow. In this case, besides the variable $x$, there are two other position variables that appear in the equation for linear stability. Its solution yields the structure of the mode in the two directions
perpendicular to the mean fluid velocity. The perturbation with all its
dependence with respect to the coordinates is found by multiplying the
amplitude by the spatial profile of the mode. We shall not write explicitly this full dependence of the amplitude in order to keep expressions as simple as possible.  

The discussion above does not exhaust all possibilities: in some parallel flows, there are two (instead of one) coordinates with an unbounded extension: for instance in plane Poiseuille or Couette or in Blasius boundary layer. Squire' theorem states that the most unstable mode, if it exists at finite Reynolds number, is for a perturbation depending only on the stream-wise coordinate $x$. However as has been often observed, for instance in Emmons spots growing in a Blasius boundary layer, the nonlinear waves inside the spots show modulation in the span-wise direction, not along the flow. This is well compatible with the present considerations because we deal with finite amplitude perturbations, not tied by Squire' theorem. The same seems to be true for plane Poiseuille where the turbulent domains show waves with a span-wise wavenumber.  Indeed such finite amplitude wave-like perturbations are to be found numerically. 

For an infinitely extended flow in the $x$ direction it does not make physical sense to compare infinitely extended perturbations along $x$ because any initial perturbation will have a finite extent along $x$. Therefore one has to look at the dynamics of perturbations depending both on time and on  $x$, something requesting an extension of Landau' theory. This extension of the amplitude equation to space dependent perturbation is formally straightforward: consider the linear part of the equation, namely 
 \begin{equation}
\frac{d A }{dt}  =    (\delta_R + i \delta_I ) A
\mathrm{.}
\label{eq:LandQredlin}
\end{equation}
The right-hand side is the result of a standard stability analysis for small amplitude fluctuations. The quantity $\delta_I $ is the pulsation of this fluctuation, and $\delta_R $, if positive in the unstable domain and negative otherwise. We consider now what happens at the threshold of instability if the fluctuation under consideration is not exactly at the critical wave number $Q$ where the rate of growth is zero. If the wave number is slightly different, the rate of growth is lessened: otherwise the rate of growth would be positive for a wave number slightly off the critical $Q$. If one takes a $x$-dependent amplitude, this decrease of the rate of growth as the wave number differs from $Q$ is represented by the addition of a term  $ (\delta_R + i \delta_I )  \frac{\partial^2 A}{\partial x^2}$ to the right-hand side of Eq.(\ref{eq:LandQredlin}), with $\delta_R $ positive in order to bring a negative contribution to the rate of growth for a wave number slightly different of $Q$. Moreover the coefficient $ i \delta_I$, pure imaginary, is for the shift of the pulsation  $\delta_I$ of the oscillations when their wavenumber is not exactly $Q$. There is also an unwritten part linear in the derivative $ \frac{\partial A}{\partial x}$, which can be absorbed by a Galilean transform. Therefore the linear part of the amplitude equation near threshold, including a possible $x$-dependence, reads: 
 \begin{equation}
\frac{\partial A }{\partial t} =    (\delta_R + i \delta_I ) A + (\alpha_R + i \alpha_I )  \frac{\partial^2 A}{\partial x^2}
\mathrm{.}
\label{eq:LandQredlin1}
\end{equation}
The full amplitude equation we shall study is derived by adding to this linear equation the same nonlinear part as in Eq.(\ref{eq:LandQred}) 

 \begin{equation}
\frac{\partial A }{\partial t}  =    \delta_R A + (\beta_R + i \beta_I) |A|^2 A - (\gamma_R + i \gamma_I ) |A|^4 A + (\alpha_{R} + i \alpha_{I})  \frac{\partial^2 A}{\partial x^2} 
\mathrm{.}
\label{eq:LandQfull1}
\end{equation}

Note that one may get rid of two parameters among the whole set of the quintic CGL equation, for example by using the following change of variables 
$t \to  \hat{t} = \frac{\gamma_{R}}{\beta_{R}^{2}} t$,  $A \to \hat{A}= \sqrt {\frac{\beta_{R}}{\gamma_{R}} } A$,  and
$x \to \hat{x} = \frac{\beta_{R}}{(\alpha_{R} \gamma_{R})^{1/2}} x$.

Dropping the hat superscripts it gives the quintic CGL equation we shall consider below
 \begin{equation}
\frac{d A }{dt}  =  \delta A  + ( 1+ i \beta) |A|^2 A - (1 + i \gamma) |A|^4 A +(1+ i \alpha)  \frac{\partial^2 A}{\partial x^2} 
\mathrm{,}
\label{eq:scaledLand}
\end{equation}

where $ \delta=\frac{\delta_{R} \gamma_{R}}{\beta_{R}^{2}}$, $\beta=\frac{\beta_{I}}{\beta_{R}}$ and  $\gamma=\frac{\gamma_{I}}{\gamma_{R}}$.  For the scaled equation (\ref{eq:scaledLand}) the condition (\ref{eq:cond1})  for subcritical behavior writes $\delta > -\frac{1}{4}$.

This type of amplitude equation is sometimes called Ginzburg-Landau equation. Besides the fact that it is a priori more general than the reaction-diffusion type of equation introduced for subcritical bifurcation in flows, it has another important property not shared by reaction-diffusion models. The reaction-diffusion case is where the imaginary coefficients are absent. This is what happens when the instability yields a steady pattern like in Rayleigh-B\'enard or B\'enard-Marangoni instability. But, in the case of disturbances of parallel flows this is no more true and imaginary terms are present, because unstable perturbations are wave-like. A consequence of the existence of imaginary terms is the vanishing of the gradient flow property: the evolution of the system does not tend anymore to minimize a real functional. This implies in particular that the stability of solutions of the intensity equation Eq.(\ref{eq:LandQredint}) cannot be extended to solution of Eq.(\ref{eq:LandQredlin1}). Therefore a plane wave solution which is stable for perturbations of the {\it{intensity}} can be unstable for perturbations of the {\it{phase}}, something obviously impossible for the reaction-diffusion case. This phase instability belongs to the class of the Benjamin-Feir instability and is analysed as follows. To lighten this analysis let us write Eq. (\ref{eq:LandQfull1}) in the condensed form:
\begin{equation}
\frac{\partial A }{\partial t}  =    A f(|A|^2)+  i A g(|A|^2) + (1 + i \alpha )  \frac{\partial^2 A}{\partial x^2}
\mathrm{,}
\label{eq:LandQfull2}
\end{equation}
where the function $f(.)$ and $g(.)$ are real function of the positive variable $|A|^2$. 

Without the diffusion term, the relation  we get the homogeneous plane-wave solution which obeys the relation $A^{*} \frac{\partial A }{\partial t}  +c.c.=0$, that gives
\begin{equation}
 A _{h}(t)= \mathcal{A} e^{i\phi_{0}} e^{i g_{0} t }
 \mathrm{,}
\label{eq:planewave}
\end{equation}
where $ \mathcal{A} $ is the modulus of the complex amplitude, a positive quantity,  $g_{0}=g(\mathcal{A}^2)$  and $\phi_{0}$ an arbitrary phase.

\subsubsection{Benjamin-Feir instability}
In order to derive the stability condition of the nontrivial solution \ref{eq:planewave}  we consider small perturbations of the amplitude and phase, of the form
\begin{equation}
 A(x,t)= \mathcal{A} (1+ \alpha^{2} s(x,t)) e^{i\phi_{0} +i g_{0}t } e^{i \alpha \eta (x,t)} 
 \mathrm{,}
\label{eq:1}
\end{equation}

where the $\alpha$  factors in front of $s$ and $\eta$ are put for convenience. Including this form in equation (\ref{eq:LandQfull2}) and separating the real and imaginary parts we get
the two coupled equations for the fluctuations of the amplitude and phase respectively, 
\begin{equation}
 \partial_{t}  s= s''-\eta''+2 c_{f} s  + \mathcal{N}_{s}
 \mathrm{,}
\label{eq:dts}
\end{equation}

and 
\begin{equation}
 \partial_{ t} \eta= \alpha^{2 }s'' +\eta'' +2 c_{g}s  + \mathcal{N}_{\eta}
 \mathrm{,}
\label{eq:dteta}
\end{equation}
 where  a prime symbol means the x-derivative and the coefficients $c_{f}=f^{(1)}_{0}\mathcal{A}^2$ , $c_{g}= \alpha g^{(1)}_{0}\mathcal{A}^2$  came  from Mac Laurin expansion of $f(\mathcal{A}^2)$ and $g(\mathcal{A}^2)$ close to the homogeneous solution,  with $f^{(n)}_{0}= [\frac{d^{n}f(|A|^2)}{d|A|^{2n}} ]_{|A|=\mathcal{A}}$  (similar for $g$) and $\mathcal{N}_{s}$ and  $\mathcal{N}_{\eta}$  stands for the nonlinear parts,
 
\begin{equation}
\mathcal{N}_{s}= -\eta'^{2}(1+\alpha^{2} s) - \alpha^{2 }\eta''s + s^{2} 2\alpha^{2}( f^{(1)}_{0}+f^{(2)}_{0})\mathcal{A}^2
 \mathrm{,}
\label{eq:Ns}
\end{equation}
\begin{equation}
\mathcal{N}_{\eta}= - \frac{s''}{1+\alpha^{2}s} - \alpha^{2} \eta'^{2} + \frac{2\alpha^{2}}{1+\alpha^{2}s}\eta' s_{x}+ \alpha^{2} s^{2} (\alpha g^{(1)}_{0}+2g^{(2)}_{0})\mathcal{A}^2
 \mathrm{.}
\label{eq:Neta}
\end{equation}

The coupled equations (\ref{eq:dts})-(\ref{eq:dteta})  are written in the form $ \partial_{ t} \mathbf{u}=   \mathcal{L} \mathbf{u} + \mathcal{N}[\mathbf{u}]$ where the vector $\mathbf{u}$ is defined by $\mathbf{u^{\dagger}}=(s,\eta)$   and the linear operator  writes

\begin{equation}
 \mathcal{L}= 
\left(
\begin{array}{ccc}
 \partial_{xx} +2c_{f}      \qquad - \partial_{xx}  \\
\alpha^{2}   \partial_{xx}+2c_{g}       \qquad \partial_{xx}   
\end{array}
\right)
 \mathrm{.}
\label{eq:linear}
\end{equation}

The linear part of the system (\ref{eq:dts})-(\ref{eq:dteta}) can be solved in the Fourier space by assuming $\mathbf{u} \propto  \exp(\lambda t -i k x)$ which gives
\begin{equation}
\lambda \mathbf{u}=
\left(
\begin{array}{ccc}
 -k^{2} +2c_{f}      \qquad k^{2}  \\
-\alpha^{2}  k^{2}+2c_{g}       \qquad    -k^{2} 
\end{array}
\right)
 \mathbf{u}
 \mathrm{.}
\label{eq:lineark}
\end{equation}

The determinant of $ \partial_{ t} \mathbf{u}-  \mathcal{L} \mathbf{u}$ vanishes for $\lambda^{2}-2(c_{f}-k^{2})\lambda+C=0$  where $C=k^{4}(1+\alpha^{2})-2k^{2}(c_{f}+c_{g})$, that gives
 $\lambda= (c_{f}-k^{2}) \pm \sqrt{(c_{f}-k^{2})^{2}-C}$.
 We are interested in small k range, where the
 rate of growth of the most unstable mode becomes
\begin{equation} 
\lambda= \frac{k^{2}(c_{f}+c_{g})-\frac{1+\alpha^{2}}{2}k^{4}}{(-c_{f})  + k^{2}}
 \mathrm{,}
\label{eq:lyap}
\end{equation}
where we recall that $(-c_{f})$ is negative along the solid line of Fig.\ref{fig:subcr}.
It follows that  any long waves ( small k  values)  are unstable if  $c_{f}+c_{g}>0$, or
 \begin{equation} 
f^{(1)}_{0}+ \alpha g^{(1)}_{0}>0
 \mathrm{,}
\label{eq:BF}
\end{equation}
 
 which defines the domain of the  Benjamin-Feir instability.
Note that the relation (\ref{eq:BF})  reduces to the usual relation $1+\alpha \beta <0$ in the well-known  case of CGL equation with cubic nonlinearity written in the form 
$ \partial_{t} A =     (1+ i \alpha) \partial_{xx}A+ A  - (1+i \beta) |A|^2 A$ where one has $f^{(1)}_{0}=-1$ and $g^{(1)}_{0}=-\beta$.
For the quintic CGL equation ( \ref{eq:scaledLand}), we have 
\begin{equation} 
f(I)= \delta +I-I^{2}
 \mathrm{,}
\label{eq:f}
\end{equation}

and 
\begin{equation} 
g(I)= \beta I-\gamma I^{2}
 \mathrm{,}
\label{eq:g}
\end{equation}

that gives $f^{(1)}_{0}(I)=1-2I(\delta) $ and $g^{(1)}_{0}  =\beta- 2\gamma I(\delta)$ where $I(\delta)$ is the stationary solution of the Landau equation,

 $$I(\delta)= |A|^2 = (1+\sqrt{1+4\delta})/2$$
  Inserting this expression into the two previous ones, we find that the unstable Benjamin-Feir  domain is for
 \begin{equation} 
I(\delta) < \frac{1+\alpha \beta}{2(1+\alpha \gamma)}   
 \mathrm{,}
\label{eq:BFq}
\end{equation}
if the denominator of the r.h.s. is positive, or $I(\delta) > \frac{1+\alpha \beta}{2(1+\alpha \gamma)}$ in the opposite case.

 In the particular case of  CGL with a  real  coefficient for the quintic term ($\gamma=0$), Benjamin-Feir instability occurs when the two conditions are fulfilled
 \begin{equation} 
\alpha \beta > 0  \qquad
-\frac{1}{4}  < \delta <-\frac{1}{4} + \frac{\alpha^{2} \beta^{2}}{4}
 \mathrm{.}
\label{eq:bf0}
\end{equation}

\subsection{Phase equation}
Here we consider the case $$f^{(1)}_{0}+ \alpha g^{(1)}_{0}= \epsilon^{2}$$ with $\epsilon$  small enough to describe the  close vicinity of Benjamin-Feir threshold where phase turbulence may occur. This conjecture is based upon previous studies devoted to the CGL (cubic case) \cite{lega}-\cite{chate} where the fluctuations of the phase has been observed  numerically (with no defects close to the threshold) and described by Kuramoto-Sivashinsky (K-S) equation , which extends a  property already derived by Kuramoto for coupled oscillators \cite{Kuramoto}.   Numerical simulations of the Landau-Ginzburg equation on finite domains indicate that close to the Benjamin-Feir threshold,  defining $1+\alpha \beta=-\epsilon^{2}$,  the fluctuations of the phase are of order $\epsilon^{2}$ around the global phase $\phi_{0}$, while the fluctuations of the amplitude remains much smaller, of order $\epsilon^{4}$ \cite{num-miguel}. The persistence of phase turbulence on infinite domains is not yet proved, while its existence on finite domains has  been stated \cite{gvb} rigorously.
Here we use same notations as \cite{gvb} and derive K-S equation for the phase dynamics of quintic CGL .

Let consider the two coupled equations (\ref{eq:dts})-(\ref{eq:dteta}) and assume that the phase fluctuations are one order of magnitude larger than the amplitude fluctuations and that this occurs on large spatial and temporal scales. More precisely we set
 $s \sim \epsilon^{4}$ , $\eta \sim \epsilon^{2}$, $t\sim \epsilon^{-4}$, $x \sim \epsilon^{-1}$.  Expanding equations  (\ref{eq:dts})-(\ref{eq:dteta}) in terms of $\epsilon^{n}$ we find that the balance  requires to cancel the  lowest order terms, of order $\epsilon^{4}$, that gives for the amplitude equation the solvability condition
 
  \begin{equation} 
s''  - \eta'' =- 2c_{f}s+( \eta')^{2}
 \mathrm{,}
\label{eq:s4}
\end{equation}

This relation means that  the l.h.s. of equation (\ref{eq:s4}) is actually of order 6. The amplitude $s(x,t)$  can be expressed  as 
  \begin{equation} 
s(\eta) =- \frac{1}{2}G( \eta'' +\eta'^{2}) 
 \mathrm{,}
\label{eq:s4b}
\end{equation}

where  $G(.) $ is the convolution operator  with the function $\mathcal(G)(x)= -\frac{1}{c_{f}+\frac{\partial{xx}}{2}}$, or in the Fourier space $\tilde{G}$ acts multicatively,  with $\tilde{G}=\frac{1}{-c_{f}+\frac{k^{2}}{2}}$.

Introducing this relation into equation \ref{eq:dteta} gives the sixth order balanced equation
  \begin{equation} 
\partial_{t} \eta=- \frac{1+\alpha^{2}}{2} G( \eta'''') -\epsilon^{2} G(\eta'') -(1+\alpha^{2}) (\eta')^{2}  
 \mathrm{,}
\label{eq:eta6}
\end{equation}
from
which we recover the Benjamin-Feir criterion (\ref{eq:BF}). More precisely linear stability analysis in Fourier space gives the dispersion relation (\ref{eq:lyap})  which shows that there are linearly unstable modes for $k < \epsilon <<1$ growing at most like $\exp{\epsilon^{4} t}$. Then one may infer that the  dynamics of equation (\ref{eq:eta6}) is dominated by the modes belonging to the small $k$ region, the high $k$ modes being slaved to them. For $k<<1$ we have $G \approx -\frac{1}{c_{f}}$, that gives the Kuramoto-Sivashinsky equation
  \begin{equation} 
(-c_{f})\partial_{t} \eta=- \frac{1+\alpha^{2}}{2} \eta'''' -\epsilon^{2} \eta'' -\frac{1+\alpha^{2}}{-c_{f}} (\eta')^{2}  
 \mathrm{,}
\label{eq:KS}
\end{equation}
 
A more complete derivation should be done  to ensure that next order terms remain bounded. This can be made following the lines of ref.\cite{gvb}

\section{Conclusion and perspectives}

To conclude we have shown that a complex amplitude equation is able to
describe weakly subcritical bifurcations which lead to non trivial space
and time dependent solutions.  We have found that the plane wave solutions
of this amplitude equation may be phase-stable or unstable with respect to
the Benjamin-Feir instability. Above (and very close to) this instability
threshold, the  fluctuations of the intensity are slaved to the  phase
fluctuations
and remains much smaller. 
This describes the bifurcation to turbulence in parallel flows without the addition of a somewhat arbitrary noise to take into account the turbulent character of the bifurcated phase. In principle such an amplitude equation can be derived from the Navier-Stokes equations, although this is quite cumbersome practically. For instance to yield quantitative predictions one needs to be at a weakly subcritical transition. In the case of a pipe flow this implies that  one has two adjustable parameters at one's disposal to make small both the rate of instability and the real part of Landau coefficient (parameter $\delta$ above) in the amplitude expansion. One such parameter is obviously the flow velocity, the other one could be found by exploring a range of possible cross sections of the pipe.  

Our presentation outlined how nontrivial effects of nonlinearity may change qualitatively what is observed in unstable flows and perhaps in other examples of instabilities in continuous media: somehow linear stability of the base solution has often little to do with what is observed as, for instance, the occurrence of localized turbulent spots in parallel flows. Their explanation relies on a chain of arguments all based on properties of the nonlinear amplitude equation, which connects the subject to the first impetus given by Landau's paper of 1944  \cite{landau}. The extension of Landau's idea to space dependent perturbation, including with the nonlinear terms, leads quite naturally to the idea that perturbations grow not only by linear instability  but also by contamination of neighboring domains with a turbulent inside. We outlined the most salient feature of the theory of this effect of contamination, including the possibility that the "turbulent" domain may be or not in an actually turbulent state depending if it is Benjamin-Feir stable or unstable. 

{}
 \ifx\mainismaster\UnDef

\thebibliography{99}
\bibitem{gradflow} We mean by gradient flow a dynamical system such that its equations of motion can be written as $$\frac{d J}{dt}  =  - \frac{\partial V}{\partial J} \mathrm{,}$$ where $V(.)$ is a real quantity and $J$ either a set  of real variables or even a function of space. In the latter case, the derivative  $\frac{\partial V}{\partial J} $ becomes a Fr\'echet derivative and $V$ a functional including space derivatives of $J$. Such gradient flow (no confusion should come between the flowintroduced now and the flow described by the fluid dynamics equations) have the property that the evolution tends to lower the energy $V$, including in the case where this functional includes a square gradient to yield a diffusion term in the dynamical equation.  If there are two steady solution, namely if the energy function $V$ has two uniform states minimizing locally (in the space of constant values of $J$) the energy, the gradient flow dynamics will tend to lower as much as possible the value of $V$. Unless special initial conditions are chosen (for instance an uniform $J$ at a metastable value) the evolution will tend to bring down the energy and to replace everywhere $J$ by its value minimizing $V$. With more or less random initial conditions this means that the evolution tends to an uniform final state with the value of $J$ bringing $V$ to its lowest value. This final state is uniform in space in order to bring to zero the square gradient in $V$ with a positive coefficient. The Navier-Stokes equation cannot be written as a gradient flow. Therefore it may be seen as slightly abusive to model fluid dynamics by equations having this property of being a gradient flow. In this sense, the amplitude equation (Eq.(\ref{eq:LandQfull2})) makes a fair candidate for representing fluid dynamics because it has the gradient flow property only when the imaginary terms are zero. In particular, when the finite amplitude waves are Benjamin-Feir unstable, this equation describes sustained turbulent solution, something obviously impossible in gradient flow dynamics.
\endthebibliography
 \end{document}
  \fi